\documentstyle[epsf]{mn}

\def\VEV#1{\left\langle #1\right\rangle}

\long\def\comment#1{}
\def\VEV#1{\left\langle #1\right\rangle}
\def\fun#1#2{\lower3.6pt\vbox{\baselineskip0pt\lineskip.9pt
  \ialign{$\mathsurround=0pt#1\hfil##\hfil$\crcr#2\crcr\sim\crcr}}}

\newlength{\tskip}
\newlength{\colwidth}

\begin{document}


\title{Intrinsic and Extrinsic Galaxy Alignment}

\author[P. Catelan, M. Kamionkowski, R. D. Blandford]{Paolo
Catelan, Marc Kamionkowski and Roger D. Blandford\\
California Institute of Technology, Mail Code 130-33, Pasadena,
CA 91125, USA}

\date{May 2000}

\maketitle

\begin{abstract}
We show with analytic models that the assumption of
uncorrelated intrinsic ellipticities of target sources that is
usually made in searches for weak gravitational lensing due to
large-scale mass inhomogeneities (``field lensing'') 
is unwarranted.  If the orientation of the galaxy image is
determined either by the angular momentum or the shape of the
halo in which it forms, then the image should be aligned
preferentially with the component of the tidal gravitational
field perpendicular to the line of sight.  Long-range
correlations in the tidal field will thus lead to long-range
ellipticity-ellipticity correlations that mimic the shear
correlations due to weak gravitational lensing.  We calculate
the ellipticity-ellipticity correlation expected if halo shapes
determine the observed galaxy shape, and we discuss
uncertainties (which are still considerable) in the predicted
amplitude of this correlation.  The ellipticity-ellipticity
correlation induced by
angular momenta should be smaller.  We consider several methods for
discriminating between the weak-lensing (extrinsic) and intrinsic
correlations, including the use of redshift information.  An
ellipticity--tidal-field correlation also implies the existence
of an alignment of images of galaxies near clusters.  Although
the intrinsic alignment may complicate the interpretation of
field-lensing results, it is inherently interesting as it may
shed light on galaxy formation as well as on structure formation.
\end{abstract}


\maketitle

\section{INTRODUCTION}

Searches for weak gravitational lensing due to large-scale mass
inhomogeneities are coming of age.  Ellipticities of
high-redshift sources are taken to be indicators of the shear
field induced by weak gravitational lensing by mass
inhomogeneities along the line of sight, and shear-shear
correlations can be used as a probe of the lensing-mass
distribution
\cite{gun67,mir91,bla91,kai92,bar92,bar99}.
The advantage of weak lensing is that it determines the power
spectrum (as well as higher-order statistics; e.g., Bernardeau,
van Waerbeke \& Mellier 1997; Munshi \&
Jain 2000; Cooray \& Hu 2000) for the {\it mass} rather
than the light.  In just the past few months, four groups have
reported the first detections of such ``field lensing''
\cite{BacRefEll00,KaiWilLup00,Witetal00,Waeetal00}.

Noise for the weak-lensing signal is provided by the intrinsic
ellipticities of the sources.  With a sufficiently large sample
of sources, the random orientation of these sources can be
overcome.  One thus looks for an ellipticity correlation in
excess of the Poisson noise provided by randomly oriented
intrinsic ellipticities.  The analysis always assumes that the
intrinsic orientations of the sources are entirely random and
isotropically distributed.  The point of this paper will be to
demonstrate that this should not be the case.

To do so, we consider two {\it ansatzen} for the origin of the
ellipticity of the high-redshift sources.  We first suppose that
the ellipticity of the galaxy image may be determined primarily
by the shape of the halo in which it forms; this might be
expected if the sources are isolated ellipticals. In this case,
a modification of the spherical-top-hat model for gravitational
collapse in a tidal field suggests a preferential
elongation of the galaxies along the direction of the tidal
field.  We show that in this case, long-range correlations in
the ellipticities of widely-separated sources are proportional
to long-range correlations in the tidal field, and thus to
correlations in the mass distribution.

We then consider what happens if the orientation of the image is determined 
by the angular momentum of the halo in which it forms; this
should be a good description if the sources are disk galaxies.
The simplest hypothesis---adopted in nearly all
disk-formation models (e.g., Dalcanton, Spergel \& Summers 1997; 
Mo, Mao \& White 1998; Buchalter et al. 2000)---is
that the plane of
the disk is perpendicular to the angular-momentum vector of the
galactic halo in which the disk forms.  According to linear
perturbation theory, a galactic halo acquires its angular
momentum via torquing of the aspherical protogalaxy in the
tidal gravitational field that arises from the large-scale mass
distribution
\cite{Hoy49,Pee69,Dor70,Whi84,HeaPea88,CatThe96}.
Averaging over all possible orientations of the protogalaxy, the
disk orientations are correlated with the tidal field.  In this case,
long-range correlations in the ellipticities are expected to be
smaller, as they will be proportional at lowest order to the
square of the correlations in the tidal-field and/or mass
distribution.

In the next Section, we review briefly the statistics used
to describe the weak-lensing signal.  In Section 3, we explan
how ellipticals should be preferentially elongated along the
direction of the tidal field, and we present the
calculation of the shear power spectrum for this case.
Section 4 presents numerical results.  In Section 5, we show how
tidal torquing can align galaxies preferentially along the tidal
gravitational field, and we explain why this should lead to
smaller ellipticity correlations that are of higher order in the 
mass correlation.
In Section 6 we put forth some ideas for
disentangling the intrinsic and weak-lensing signals, including
the use of redshift information, and we predict a corresponding
alignment in the images of galaxies near clusters.  We close
with some concluding remarks in Section 7.

During preparation of this paper, we learned of related work
(using numerical simulations) by Heavens, Refregier \& Heymans
(2000) and Croft \& Metzler (2000).  Our analytic approach
should complement their numerical work and perhaps help shed
some light on the origin of their observed correlations.  The
analytic calculation should also be useful in determining the
correlation at large angular separations, where it becomes
increasingly difficult to measure in simulations.  Our analytic
approach also suggests some possible
intrinsic/weak-lensing discriminators.

\section{FORMALISM AND STATISTICS}

As discussed, e.g., in Kamionkowski et al. (1998), weak lensing
induces a stretching of images on the sky at position
$\vec\theta=(\theta_y,\theta_z)$ described by $\epsilon_+$, the
stretching in the $\hat\theta_y - \hat\theta_z$ directions, and
$\epsilon_\times$, the stretching along axes rotated by
$45^\circ$ (we take the line of sight to be the ${\bf \hat
x}$ direction).  Possible definitions for these ``ellipticities''
are discussed, e.g., by Blandford et al. (1991), and strategies
for averaging over the redshift distribution of the sources are
described by Kaiser (1992).  The
weak-lensing shear, $\gamma(\vec\theta)$, can be recovered
through its Fourier transform \cite{Ste96},
\begin{equation}
     \tilde \gamma(\vec\ell) = { (\ell_y^2 -\ell_z^2) \tilde
     \epsilon_+(\vec\ell) + 2\ell_y \ell_z \tilde
     \epsilon_\times(\vec\ell)  \over
     \ell_y^2 + \ell_z^2},
\label{eqn:gamma}
\end{equation}
where $\tilde\epsilon_+(\vec\ell) = \int\, \vec\theta \,
\epsilon_+(\vec\theta) {\rm e}^{i \vec\ell \cdot \vec\theta},$
is the Fourier transform of the ellipticity (and similarly for
the other quantities).
Statistical homogeneity and isotropy guarantee that the Fourier
coefficients, $\tilde\gamma(\vec\ell)$, have zero mean and variances,
\begin{equation}
    \VEV{ \tilde\gamma^*(\vec\ell)
    \tilde\gamma(\vec\ell')} = (2\pi)^2
    \delta(\vec\ell -\vec\ell') C(\ell),
\end{equation}
where $C(\ell)$ is the weak-lensing power spectrum.  The mean-square
ellipticity (which is usually taken to be the mean-square shear) 
smoothed over some circular window of radius $\theta_p$ is
\begin{equation}
     \VEV{ (\gamma_g^s)^2} = \int \, {d^2 \vec\ell \over
     (2\pi)^2} \, C(\ell)\, |\widetilde W(\vec\ell)|^2,
\label{eq:smoothedvariance}
\end{equation}
where $\widetilde W(\ell) = 2 J_1(x)/x$ is the Fourier-space
window function, $J_1(x)$ is a Bessel function, and $x \equiv
\ell\theta_p/\sqrt{\pi}$ (e.g., Mould et al. 1994).

Two nonzero 2-point ellipticity-ellipticity correlation functions
can be constructed from the two ellipticities, $\epsilon_+^r$
and $\epsilon_\times^r$,
determined by taking the coordinate axes to be aligned with the
line connecting the two correlated points.
These correlation functions are
$C_1(\theta)=\VEV{\epsilon_+^r(\vec\theta_0)
\epsilon_+^r(\vec\theta_0+\vec\theta)}$ and
$C_2(\theta)=\VEV{\epsilon_\times^r(\vec\theta_0)
\epsilon_\times^r(\vec\theta_0+\vec\theta)}$, and
they are given in terms of the power spectra by
\begin{eqnarray}
     C_1(\theta)+C_2(\theta) &=& \int_0^\infty {\ell\, d\ell \over 2\pi}
     C(\ell) J_0(\ell\theta), \nonumber
      \\
     C_1(\theta)-C_2(\theta) &=& \int_0^\infty {\ell\, d\ell \over 2\pi}
     C(\ell) J_4(\ell\theta),
\end{eqnarray}
where $J_\nu(x)$ are Bessel functions.  These are the
correlation functions that have been found to be nonzero.  The
power spectrum can be recovered from these correlation functions 
through an inverse transform (see, e.g., Kamionkowski et al. 1998).

\section{HALO SHAPE DISTORTIONS}

\begin{figure}
\centerline{\epsfxsize=\colwidth\epsffile{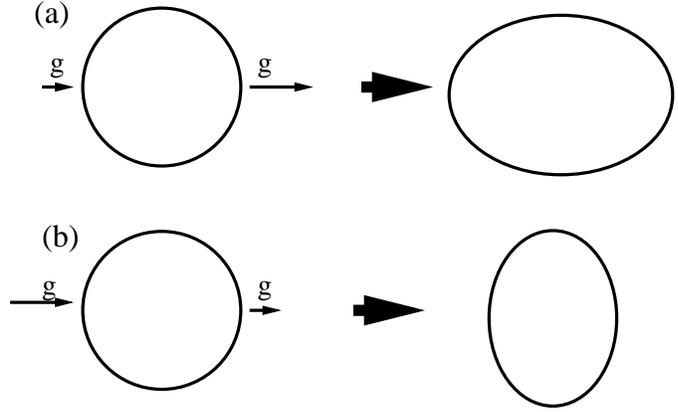}}
\caption{\label{fig:ellipse}
     The two panels show how a tidal field can either
     elongate or compress a galactic halo.  The arrows labeled
     ``g'' indicate the size and direction of the gravitational
     field.  In (a) the tidal field stretches the halo, and in
     (b) the tidal field compresses the halo.}
\end{figure}

The observed shape of a galaxy may be determined, at least in part, by the
shape of its halo.  Suppose a spherical overdensity undergoes
gravitational collapse to form a galaxy in a region of constant
tidal gravitational field.  Then the acceleration on one side of
the galaxy will differ from that at the other side, and the
resulting gravitational collapse, which would have otherwise
been spherically symmetric, will be anisotropic.  As illustrated 
in Fig. \ref{fig:ellipse}, the resulting
ellipticity could be either positive or negative depending on
whether the tidal field compresses the sphere (resulting in an
oblate halo) or stretches the sphere (resulting in a prolate halo).
Although overdensities are not all expected to be spherical,
there should still be some net elongation of halos that form in a 
given tidal field after averaging over all shapes.
Moreover, the shape of the luminous
galaxy should be determined, at least in part, by the shape of the
halo in which it forms.  For example, the stellar distribution of
luminous ellipticals that form in isolation should trace the
shape of the halo.  Alternatively, a gaseous pancake could form
in a plane perpendicular to the
tidal field if the tidal-field axis undergoes gravitational collapse
first.  

The symmetry dictates that the mean ellipticity of a
population of galaxies that form in a given tidal field will
be related to the potential via,
\begin{eqnarray}
     \epsilon_+ &=& C (\partial_y^2-\partial_z^2) \phi,
     \nonumber \\
     \epsilon_\times &=& 2 C \partial_y  \partial_z \phi,
\label{eqn:haloresult}
\end{eqnarray}
where $C$ is a constant to be discused further below.
To see that equation (\ref{eqn:haloresult}) is the lowest-order
term that encodes the expected qualitative behavior, consider
the motion of a sphere of test masses in a slowly-varying
(spatially) potential $\phi(\vec x)$.  If we Taylor expand
$\phi(\vec x)$ about
the origin, the zeroth order term has no physical effect.  The
linear term gives rise to a uniform translation of the sphere
(constant gravitational field $\vec g =\nabla \phi$), but does
not affect its shape.  The quadratic term (the tidal field) will
change the shape of the sphere, as it gives rise to different
accelerations at different points of the sphere, as shown in
Fig. \ref{fig:ellipse}.  To lowest order in $\phi$, the
ellipticity induced in the sphere is given by equation
(\ref{eqn:haloresult}).

We now proceed to show that long-range correlations in the tidal 
field should lead to ellipticity-ellipticity correlations that
mimic the weak-lensing shear.  The shear measured
in direction $\vec\theta$ will be the integrated ellipticity
along the line of sight,
\begin{eqnarray}
     \epsilon_+(\vec\theta) &=& \int_0^\infty \,dx\, g(x)
     \epsilon_+(x, x\theta_y,x\theta_z) \nonumber \\
     &=& C \int_0^\infty \, dx
     \, g(x) (\partial_y^2 - \partial_z^2)
     \phi(x,x\theta_y,x\theta_z),
\end{eqnarray}
where $g(x)$ is the distribution of sources along the line of
sight normalized to unity, and $x$ is the comoving
distance.  We have assumed the Universe to be flat, as the
cosmic microwave background seems to indicate (Kamionkowski,
Spergel \& Sugiyama 1994; Miller et al. 1999; de Bernardis et
al. 2000; Hanany et al. 2000). Using the generalized Limber's
equation (Kaiser 1992), as well as the definition of $\gamma$ in
equation (\ref{eqn:gamma}), we find that intrinsic alignments
yield a shear power spectrum,
\begin{equation}
     C(\ell)=C^2\left({3\over2} \Omega_0 H_0^2\right)^2 
     \int\, dx \, {g^2(x) \over x^2} \, P(\ell/x).
\label{eqn:firstintegral}
\end{equation}
If the ellipticity-ellipticity correlation evolves with time, then the
time dependence of the mass power spectrum should be taken into
account in the integrand.  However, we are supposing that the
ellipticity correlations are fixed by some primordial density
field, so the time evolution does not matter.  The overall
normalization of $P(k)$ also does not matter, as the
normalization of the shear power spectrum will be
fixed below to match the observed source ellipticities.

In principle, a complete galactosynthesis model would allow us
to tell how the ellipticity of the luminous galaxy is
related to the shape of the dark halo in which it forms, but in
practice, we are far from being able to do this.  We thus choose 
to empirically estimate the constant of proportionality $C$
between the ellipticity and the tidal field.

To do so, we make the {\it ansatz} that the halo shape is the
only factor that determines the ellipticity of the observed
luminous galaxy.  We can then use our model to 
calculate the expected rms ellipticity of individual galaxies
and compare this prediction with the typical
source ellipticity to fix $C$.  We use $\bar\epsilon \simeq
0.15$ (M. Metzger, private communication), although it
might be larger for some high-redshift populations.  
Doing so, we find
\begin{eqnarray}
     \VEV{\epsilon^2}&=&\VEV{\epsilon_+^2+\epsilon_\times^2}\nonumber \\
     &=&{8C^2 \over 15} \VEV{\left( \nabla^2 \phi \right)^2}
     \nonumber \\
     &=& {8C^2 \over 15} \int\, {d^3 {\bf k} \over (2\pi)^3}
     k^4 P_\phi(k),
\end{eqnarray}
where $P_\phi(k)$ is the power spectrum for the gravitational
potential which is related to the mass power spectrum $P(k)$,
through the Poisson equation.  We thus obtain
\begin{equation}
     \VEV{\epsilon^2}= {8 \over 15} C^2 \left( {3\over2}
     \Omega_0 H_0^2 \right)^2 {1\over 2\pi^2} \int \, k^2 \,
     dk\, P(k) \left[ {3j_1(kR) \over kR} \right]^2.
\end{equation}
This fixes the constant of proportionality $C$ when we adopt
$\VEV{\epsilon^2}=\bar\epsilon^2=(0.15)^2$.  We have inserted
a top-hat window function and choose to smooth
over a radius $R=1\,{\rm Mpc}\,h^{-1}$, a characteristic scale over 
which a galaxy forms (tidal fields on smaller scales should not
contribute to the torque).  Below we will discuss how uncertainty
in the smoothing scale will affect our final results.

Changing the variable of integration in equation
(\ref{eqn:firstintegral}) from $x$ to $k$ and inserting our
expression for $C$, we get the intrinsic shear power spectrum,
\begin{equation}
     C(\ell) = { {15 \pi^2 \bar\epsilon^2 \over 4} {1\over
     \ell}} {\int \, dk\,
     g^2(\ell/k) \, P(k)  \over \int k^2 \, dk \, P(k) \left[
     {3j_1(kR) \over kR} \right]^2}.
\label{eqn:power}
\end{equation}
Thus, if the observed shape of a galaxy is determined, even
partially, by the tidal field in which it forms, then long-range
correlations in the tidal field will lead to long-range
correlations in the ellipticity, or equivalently, in the induced 
shear.  Equation (\ref{eqn:power}) bears some resemblance to the 
power spectrum for weak
gravitational lensing.  The difference is that the $g(x)$ here
is replaced by the distribution of lenses along the line of
sight, and the normalization differs.  All this should come as no
surprise, since the shear produced by weak lensing also depends
on the perpendicular components of the tidal field.  Thus,
barring the small difference expected from the different
line-of-sight distribution, the angular dependence of the
intrinsic correlation function should look quite a bit like that 
for weak lensing.

\section{NUMERICAL RESULTS}

\begin{figure}
\centerline{\epsfxsize=\colwidth\epsffile{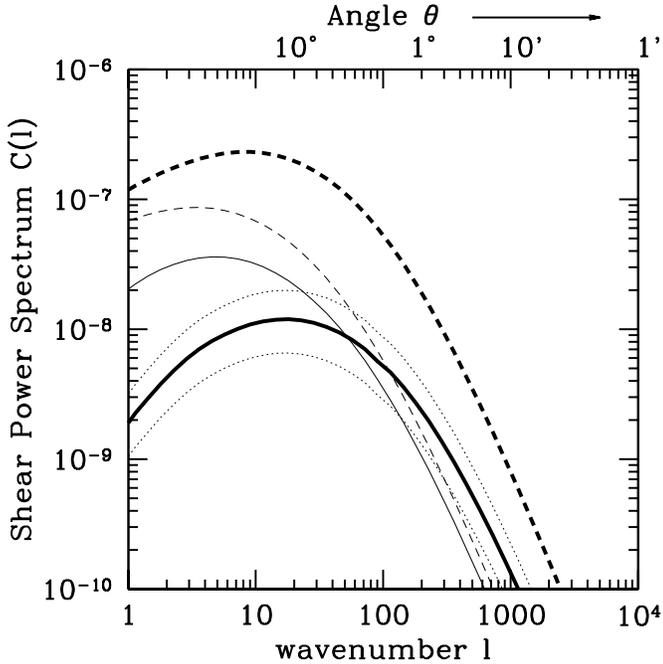}}
\caption{\label{fig:power}
     The angular shear power spectra for weak lensing (dashed
     curves) and for intrinsic alignments (solid curves) for
     a high-redshift source population with median redshift
     $z_m\sim1$ (heavier curves) and a lower-redshift source
     population with $z_m\sim0.3$ (lighter curves).  
     The upper and lower dotted
     curves show the intrinsic power spectrum that would be
     obtained (for the high-redshift population) using a
     smoothing radius of $R=2\,h^{-1}$ Mpc
     or $R=0.5\,h^{-1}$ Mpc, respectively, instead of the
     nominal value of $R=1\,h^{-1}$.  The mean source
     ellipticity is assumed here to be $\bar\epsilon=0.15$, and
     the amplitude of the intrinsic power spectrum scales with
     $\bar\epsilon^2$.}
\end{figure}

Fig. \ref{fig:power} shows results for the shear power spectrum
induced by intrinsic alignments as well as the power spectrum
expected from weak lensing.  Results are shown for a
population of sources with a mean redshift $z_m \sim1$ and a
population with a mean redshift $z_m \sim 0.3$.  The
weak-lensing power spectrum is taken from Kamionkowski et
al. (1998), and is for a cluster-abundance-normalized
cold-dark-matter model.\footnote{The amplitude of the
weak-lensing power spectrum {\it does} depend on the amplitude
of $P(k)$.}  The light dotted curves
indicate how the intrinsic power spectrum (for the high-redshift 
population) would change if the smoothing scale for the
normalization to the observed ellipticities was taken to 
be $R=0.5\,h^{-1}$ Mpc (lower dotted curve)
or $R=2\,h^{-1}$ Mpc (upper dotted curve), instead of
$R=1\,h^{-1}$.

The calculation indicates that for high-redshift sources, the
intrinsic correlation is unlikely to dominate that from weak
lensing.  However, given the uncertainty in the normalization,
and the approximation inherent in our calculation, we cannot rule out the
possibility that the intrinsic correlation might be larger than
our calculation has indicated.  There is also some chance that 
the correlation could be smaller than we have predicted.  We
have normalized our prediction by assuming that the observed
ellipticities are due entirely to the halo shape.  However,
realistically, some fraction of the intrinsic ellipticity will
be due to other effects (e.g., galactic spins; see below), and the
amplitude of our predicted signal will be accordingly lowered.
However, even if the intrinsic signal is small, it will not be
zero, and will thus need to be considered for
cosmological-parameter estimation (e.g., Hu \& Tegmark 1999)
from future precise
weak-lensing maps as well as for studies of higher-order
weak-lensing statistics.  The curves in Fig. \ref{fig:power} for the
lower-redshift population suggest that intrinsic alignment will
be more important for weak-lensing searches with shallower
surveys, such as the Sloan Digital Sky Survey and/or Two-Degree
Field.

\section{TIDAL TORQUES AND IMAGE ORIENTATIONS}

\begin{figure}
\centerline{\epsfxsize=\colwidth\epsffile{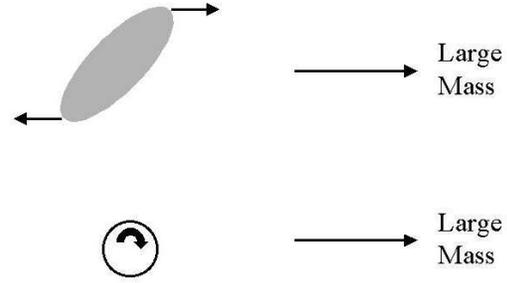}}
\caption{\label{fig:diagram}
     The upper panel shows a distant mass acting 
     upon a prolate halo and causing it to
     start spinning as shown. The torque vanishes when 
     $\theta=0,\pi/2$, cf., Eq.~\protect\ref{eqn:orientation}. 
     The lower panel shows that baryons
     in the potential well will fall in to form a disk lying in the 
     plane of the figure. When this disk is viewed 
     from the direction perpendicular to the figure, there will be no induced
     ellipticity; when viewed from the plane of the figure,
     the ellipticity will be $\sim 1$.  If the mass is replaced
     by a void, the sense of rotation is reversed, but the shape 
     remains unchanged.  Averaging over all viewing
     directions gives a non-zero mean elongation of the galaxy image along the
     projected direction of the distant mass and this will be correlated 
     among neighboring galaxies.  More generally, tidal
     gravitational fields will be induced by the large-scale
     distribution of mass (rather than by a single large mass),
     and long-range correlations in these tidal fields will
     induce long-range correlations in the ellipticities of
     widely separated galaxies.}
\end{figure}

Another possibility is that the shapes of galaxies are
determined by the angular momenta of the halos in which they
form.  This is heuristically expected to account for
the shapes of disk galaxies (while the shapes of ellipticals might
be expected to be determined more by the shape of the halo, as
discussed above).
In this Section, we will argue that the tidal-torque theory for
the origin of galactic angular momenta also suggests that galaxy
images should tend to be aligned with the component of the tidal 
gravitational field perpendicular to the line of sight.  

When the mass that will eventually collapse to form a galaxy
breaks away from the expansion, it will in general have
an anisotropic moment-of-inertia tensor ${\cal I}_{\alpha\beta}$ 
and reside in a tidal gravitational field ${\cal
D}_{\alpha\beta}=\partial_\alpha \partial_\beta \phi$
generated by the larger-scale distribution of mass, where $\phi$ 
is the gravitational potential.  The moment-of-inertia
eigenframe will most
generally not be aligned with the that of the tidal field.  A
net torque will thus be applied to the protogalaxy, and
the galactic halo that results will have an angular momentum
$L_\alpha \propto \varepsilon_{\alpha\beta\gamma} {\cal
I}_{\beta\sigma} {\cal D}_{\gamma\sigma}$, where
$\varepsilon_{\alpha\beta\gamma}$ is the Levi-Civita symbol.
(see Figure \ref{fig:diagram}).  The relation between the
ellipticity and angular momentum can ultimately be determined
empirically by fitting to the observed ellipticities (see
Section 5 below), so the absolute magnitude of ${\bf L}$ is
irrelevant.  Thus, when we refer to the ``angular momentum''
{\bf L} in the following, we will really be discussing a
vectorial spin parameter, an angular momentum that has been
scaled by the galaxy mass and binding energy to be dimensionless.

Let us assume that if a disk forms in a 
galactic halo, it will form in the plane perpendicular to the
angular-momentum vector.  If
viewed at an inclination angle $\alpha$, the ellipticity will be 
reduced by $\cos\alpha$ from the value it would have if viewed
edge on.  Taking the line of sight to be the 
${\bf \hat x}$ axis, the observed ellipticity will be
\begin{eqnarray}
     \epsilon_+ &=& f(L,L_x) (L_y^2-L_z^2) \nonumber \\
     \epsilon_\times &=& 2 f(L,L_x) L_y L_z,
\label{eqn:epsilonandL}
\end{eqnarray}
where $L=|{\bf L}|$.  

Let us now suppose that a number of galaxies form in the same
tidal field.  Since the angular
momentum (and corresponding observed ellipticity) of a given
galaxy will depend on the moment of inertia of its parent
protogalaxy, the tidal field alone does not determine the
orientation of a given disk.  
However, if the moment-of-inertia eigenframes are distributed
isotropically, a net alignment of the galaxies with the
tidal field remains after averaging over all initial moments of
inertia.

To demonstrate this, we will consider a
somewhat simplified model that should still capture the
essential physics and leave out details.
Rather than average over the entire distribution of moments of
inertia---which can be obtained, e.g., from Gaussian peak
statistics \cite{Baretal86,HeaPea88,CatThe96}---we will suppose
that each galaxy has the same eigenframe moment of inertia and
that the eigenframes are distributed isotropically.  Moreover, we will
suppose that two of the three principal moments are
equal, and the third is slightly larger.  The unequal moment
defines a symmetry axis ${\bf \hat
n}=(\sin\theta \cos\varphi,\sin\theta \sin\varphi, \cos
\theta)$, and the contributing part of the moment-of-inertia tensor
is ${\cal I}_{\alpha\beta} \propto n_\alpha n_\beta$, where
${\bf \hat n}=(\sin\theta \cos\varphi,\sin\theta \sin\varphi,
\cos \theta)$.

To simplify our illustration, we will take $f(L)=C$, where $C$
is a constant (so the ellipticity is quadratic in the spin).
Consider now a region of fixed tidal field, which we will take to be
$D_{\alpha\beta}= A k_\alpha k_\beta$, where $A$ is a constant.
The induced angular momentum is thus ${\bf L} =A ({\bf \hat n}\cdot
{\bf k}) ({\bf \hat n} \times {\bf k})$ (up to a constant
of proportionality to be dealt with below).

If ${\bf k}$ is perpendicular to the
line of sight, say in the ${\bf \hat z}$ direction, then the
observed ellipticity of a galaxy that forms from a protogalaxy
with symmetry axis in the ${\bf \hat n}$ direction is
\begin{equation}
     \epsilon_+ = C A^2 k^4 \sin^2\theta \cos^2\theta \cos^2
     \varphi, \qquad \epsilon_\times=0.
\label{eqn:orientation}
\end{equation}
So, for example, if the symmetry axis is
aligned with ($\theta=0$) or perpendicular to ($\theta=\pi/2$)
the tidal field, there will be 
no torque.  If the symmetry axis is 
in the $x$-$z$ plane, then it will be torqued about the ${\bf \hat 
y}$ direction and the resulting image will be elongated in the
${\bf \hat z}$ direction.  If the symmetry axis is in the
$y$-$z$ plane, the galaxy will spin about the line of sight and
produce a face-on (zero-ellipticity) disk.  If the tidal field
is reversed in direction (i.e., $A \rightarrow -A$), then the
same ellipticity arises.

We now assume that the orientations of the protogalaxy eigenframe
are random.  
In principle, the location and
shapes of peaks from which the protogalaxy forms depend on all
Fourier modes in the initial density field, including those that 
determine the long-range tidal field.  Although the tidal field
may play some role in determining the halo shape, as discussed
above and in Lee \& Pen (2000a), it cannot exclusively determine
the halo shape. We thus make the usual peak-background split and 
assume that the small-scale fluctuations that play the dominant
role in determining the shape of the protogalaxy are
statistically independent of larger-scale fluctuations.  

Integrating equation
(\ref{eqn:orientation}) over all angles, we find that the
inferred shear, the mean ellipticity of galaxies formed in this
tidal gravitational field, is $\epsilon_+ = C A^2 k^4 /15$,
$\epsilon_\times=0$.  Thus, although there is no one-to-one
correspondence between the tidal gravitational field in which a
galaxy forms and the orientation of the galaxy image, {\it on
average, there will be a tendency for galaxy images to be
aligned with the major axis of the the tidal gravitational field.}

The tidal field, $D_{\alpha\beta} = A k_\alpha k_\beta$, used in 
the derivation above is not the most general tidal field.
For an arbitrary tidal field perpendicular to the line of
sight---$D_{\alpha\beta}=\partial_\alpha \partial_\beta \phi$
with $\partial_x \phi=0$---the {\it mean} ellipticity of 
sources that form will be (Kamionkowski, Mackey \& White, in
preparation),
\begin{eqnarray}
     \epsilon_+ &=& {C^2 \over 15} \left[(\partial_y^2
     \phi)^2-(\partial_z^2 \phi)^2\right],
     \nonumber \\
     \epsilon_\times &=& {-2 C^2 \over 15} (\partial_y  \partial_z
     \phi) (\partial_y^2 \phi + \partial_z^2 \phi).
\label{eqn:result}
\end{eqnarray}
Galaxies that form in a tidal field parallel to the line of 
sight will form edge-on disks, but their orientations will be
isotropic thus leading to no induced shear.  
Although the detailed form of equation (\ref{eqn:result}) would
be altered if we had taken some other form for $f(L)$ (e.g., a
constant or linear dependence of $epsilon$ on $L$ instead of a quadratic
relation), we would still have found some correlation between the
induced ellipticity and the tidal field.

If the ellipticity of sources has a quadratic dependence on the
tidal field, as in equation (\ref{eqn:result}), then
correlations in the ellipticity are higher order in tidal-field
correlations.  In this case, long-range ellipticity-ellipticity
correlations should be smaller, and their power spectrum should
be stronger at smaller scales.  Moreover, the ellipticity
correlations will have some curl component, rather than
the pure curl-free power spectrum produced by weak lensing
\cite{Ste96,Kametal98} and by the correlations considered in
Section 3.  Detailed results for this case
will be presented elsewhere (Kamionkowski, Mackey \& White, in
preparation).  We could have alternatively postulated that $f(L) 
\propto L^{-1}$; i.e., an ellipticity that is proportional to
the angular momentum and thus to the tidal field.  However, even 
in this case, the {\it correlations} would still be higher
order in the tidal-field and/or mass correlations
\cite{LeePen00a,LeePen00b,Critetal00}.  For every
mass distribution, there is another in which
overdensities are replaced by underdensities and {\it vice
versa}.  Although these mass-reversed configurations will lead
to angular momenta that differ in sign, they lead to the same
ellipticity.  In constructing the statistical ensemble
from which the ellipticity correlations are measured, the
contributions from the two angular momenta cancel, and this
prohibits any ellipticity correlation that is linear in the mass 
or tidal-field correlations.

\section{INTRINSIC VERSUS EXTRINSIC DISCRIMINATORS}

\subsection{Redshift information} 

Redshift information could be used to determine
the relative contributions of the intrinsic and weak-lensing
correlations.  As Fig. \ref{fig:power} indicates, the
weak-lensing signal is larger for more distant sources (as there 
is more line of sight for the signal to accrue), while the
intrinsic correlation should actually increase for
lower-redshift sources.  We should caution, however, that
dynamical processes might dilute intrinsic ellipticity
correlations with time; this should be amenable to
further study by numerical simulations.  It is thus plausible that
low-redshift populations show no intrinsic correlation even
though high-redshift populations do.

Another possibility would be to exploit the different dependence 
of the weak-lensing and intrinsic correlations on the redshift
distribution of the sources.  Since the weak-lensing signal
accrues along the line of sight, the ellipticity correlation
between two objects nearby on the sky but widely separated in
redshift may be significant.  On the other hand, the intrinsic
ellipticity correlation should be larger for pairs of
sources that are close in redshift.  Some indication 
of this can be seen in equation (\ref{eqn:firstintegral}).
Consider a distance distribution $g(x)$ that is a top hat
centered at $x_0$ with width $\Delta x$.  The shear power
spectrum that results is inversely proportional to $\Delta x$
(as long as $\Delta x$ is not so small that the
approximation used in deriving equation
(\ref{eqn:firstintegral}) breaks down).

Along similar lines, our model predicts the existence of
correlations between ellipticities of spatially close pairs of
galaxies.  Redshift surveys should yield a good sample of close
(in redshift as well as celestial position) pairs of galaxies
with which this correlation could be searched.

\subsection{Cuts on Intrinsic Ellipticities}  

The strength of the
shear induced by weak-lensing does not depend on the intrinsic
ellipticities of the sources.  On the other hand, our
calculation [cf. equation (\ref{eqn:power})], suggests that the
magnitude of the intrinsic correlation is proportional to the
square of the average ellipticity of the sources.  Thus, the
intrinsic correlations could be reduced by using sources with
smaller intrinsic ellipticities, or by choosing isophotes that
yield images of the smallest intrinsic ellipticity.
Alternatively, one could determine the ellipticity correlation
functions for the same population of sources, but using several
different isophotes.  Since the weak-lensing signal should not
depend on the intrinsic ellipticity while the intrinsic
correlation should, comparison between these various correlation 
functions can be used to separate the intrinsic and weak-lensing 
signals in much the same way that multifrequency
cosmic-microwave-background maps disentangle the cosmological
signal from Galactic foregrounds.

\subsection{Cross-correlation with density} 

Another possibility is
cross-correlation between the shear and 
the convergence.  In addition to distorting images, weak lensing 
will also affect the density of sources on the sky, and this
might be used to isolate the weak-lensing signal.  However,
given that the tidal field is correlated with the mass (and thus 
the galaxy) distribution, there may be a similar
cross-correlation between the density of sources and the
intrinsic orientations which should be investigated further
(Kamionkowski, Mackey \& White, in preparation).

\subsection{Intrinsic correlations around clusters}  

Our model makes a
precise prediction for the relation between the mean orientation 
of galaxies and the tidal field in which they form.  Clusters
are produced at very high-density peaks of the initial density
distribution, so they should have been surrounded at early times
by large primordial tidal fields (in the radial direction).
There should thus be an alignment of nearby galaxies {\it at
redshifts comparable to the cluster}.  The algorithms
that have been developed to look for alignment of distant
background galaxies due to weak lensing by the cluster could be
applied to galaxies near the cluster to search for an alignment
with the tidal field.  If the alignment is too
small to be detectable with the finite number of galaxies
associated with one cluster, it may be possible to ``stack''
several clusters to improve the sensitivity to this alignment.
This ought to provide an alternative and possibly superior
calibration of the intrinsic alignment.

\subsection{Morphological distinctions}

We have considered two possible mechanisms for aligning galaxy
shapes---correlations with halo angular momenta and with halo
shapes.  Heuristically, the shapes of disk galaxies are expected 
to be determined more by the spin of the halo, while the shapes
of ellipticals should be determined more by the shapes of the
halos in which they form.  These two mechanisms make
considerably different predictions for long-range ellipticity
correlations.  Thus, the weak-lensing shear could be
disentangled from the intrinsic signal by using information
about the morphology of the sources.  Specifically, we expect
the intrinsic alignment of ellipticals to have more power on
larger scales relative to smaller scales than the intrinsic
alignment of spirals.

\section{CONCLUSIONS AND DISCUSSION}

We have shown that the shapes and/or angular momenta of galactic 
halos should depend to some extent on the tidal gravitational
field in which they are produced.  Long-range correlations in
the gravitational field should thus lead to long-range
correlations in the shear inferred from images of distant
galaxies.  Although uncertainties in the relation between the
luminous-galaxy shape and the halo shape prohibit us from carrying
out a ``first principles'' calculation of the correlation, we
can estimate the magnitude of a the correlation that arises if
ellipticities are determined by halo shapes by calibrating to
the observed distribution of ellipticities.  

The amplitude of the
intrinsic power spectrum is increased (decreased) with
a larger (smaller) smoothing length $R$. Realistically, there will be
factors in addition to those that we have considered that
contribute to the observed orientation, and these could decrease 
the correlation.  One example is the halo spin.  As another
example, major mergers could affect the halo shape as well as
the orientation of the disk relative to that of the halo.  
All of these effects will tend to diminish the correlations
predicted by our model.  However, results from numerical
simulations (Heavens, Refregier \& Heymans 2000; Croft \&
Metzler 2000) seem to indicate that the correlations in the
halos have not been much diluted by these effects.
Although these simulations quantify the correlations of the
parent halos, there is still a considerable amount
of physics relating the halo shape to the shape of the luminous
galaxy that cannot yet be described properly with simulations.
Heuristically, these effects should tend to diminish the
intrinsic correlations even further.  If future theoretical work 
determines that the degradation is considerable,
then the effects we are discussing will be unimportant for
interpretation of recent field-lensing detections.  However,
even if the correlation is small, it should not be zero---we
have indeed identified realistic physical effects that should
play at least some role in aligning galaxy images.  Thus,
the physical effects we have discussed here will be important for
interpretation of future more precise weak-lensing maps, as well as for
understanding the implications of measurements of higher-order
weak-lensing statistics.  

At first, this intrinsic correlation may be seen as a nuisance
for field-lensing searches.  However, the intrinsic correlation
arises from the same long-range correlations in the density
field that give rise to the weak-lensing correlation.  Moreover, 
the galaxy-formation physics that produces spins and shapes of
galaxies is itself inherently interesting.  Thus, measurement of
these intrinsic correlations would be of fundamental
significance for structure formation and galaxy formation.

\section*{ACKNOWLEDGMENTS}

We thank D. Bacon, R. Croft, R. Ellis, A. Heavens, M. Metzger,
C. Metzler, A. Refregier, M. Santos, and M. White for
discussions, and we thank especially U.-L. Pen and
M. Zaldarriaga for identifying an error in an earlier draft.
This work was supported in part by NSF AST-9900866, AST-0096023,
NASA NAG5-8506, and DoE DE-FG03-92-ER40701.

{}

\end{document}